\documentclass[twocolumn,secnumarabic,amssymb, nobibnotes, aps, prb]{revtex4-1}
\usepackage{amsmath}
\usepackage{mathtools}

\usepackage{algorithmic}
\usepackage{graphicx}
\usepackage{textcomp}
\usepackage{xcolor}
\usepackage{caption}
\usepackage{subfigure}
\usepackage{chemformula} 
\usepackage{booktabs} 
\usepackage{bm}
\usepackage{multirow}
\AtBeginDocument{ 
\heavyrulewidth=.08em
\lightrulewidth=.05em
\cmidrulewidth=.03em
\belowrulesep=.65ex
\belowbottomsep=0pt
\aboverulesep=.4ex
\abovetopsep=0pt
\cmidrulesep=\doublerulesep
\cmidrulekern=.5em
\defaultaddspace=.5em
}

\newcommand*{\mrm}[1]{\mathrm{#1}}
\newcommand*{\mtrm}[1]{\relax\ifmmode\mrm{#1}\else{#1}\fi} 
\renewcommand*{\vec}[1]{\bm{\mrm{#1}}}
\renewcommand*{\tensor}[1]{\bm{#1}}

\newcommand*{\CrCl}{\ch{CrCl3}}
\newcommand*{\CrGeTe}{\ch{CrGeTe3}}
\newcommand*{\CrBr}{\ch{CrBr3}}
\newcommand*{\CrI}{\ch{CrI3}}
\newcommand{\eg}[1]{{\textit{e.g.,} #1}}
\newcommand{\ie}[1]{{\textit{i.e.,} #1}}
\newcommand*{\Eq}[1]{Eq.\ \eqref{#1}}
\newcommand*{\Curie}{\mtrm{C}} 
\newcommand*{\TCurie}{\relax\ifmmode{\textit{T}_{\Curie}}\else{\textit{T}\textsubscript{\Curie}}\fi} 
\newcommand*{\BZ}{\mtrm{BZ}}
\newcommand*{\DeltaNN}{$\Delta_{\rm NN}$}
\newcommand*{\kB}{k_\mrm{B}} 
\newcommand*{\kBT}{\kB T} 
\newcommand*{\nth}[1]{\relax\ifmmode{^{\text{#1}}}\else{\textsuperscript{#1}}\fi} 
\newcommand*{\Over}[1]{\frac{1}{#1}}
\newcommand*{\infrac}[2]{{\begingroup #1 \endgroup /{#2}}}

\newcommand*{\RecPrimCellVol}{v_{b}} 
\newcommand*{\Equal}{\mtrm{E}} 
\newcommand*{\Other}{\mtrm{O}} 
\newcommand{\dint}[4]{\int_{#1}^{#2}#3\,\mrm{d}#4}
\newcommand*{\vk}{\vec{k}}
\newcommand*{\ii}{\mathrm{i}}
\newcommand*{\bcdot}{\mathbin{\bm{\cdot}}}
\DeclarePairedDelimiterX{\norm}[1]{\lVert}{\rVert}{#1}

\newcommand{\fix}[1]{{\color{black} #1}}
\newcommand{\new}[1]{{\color{black} #1}}
\begin{document}
\title{Computing Curie temperature of two-dimensional ferromagnets in the presence of exchange anisotropy}
\author{Sabyasachi Tiwari*$^{1,2,3}$, Joren Vanherck*$^{5,3}$, Maarten~L.~Van~de~Put$^{1}$, William~G.~Vandenberghe$^{1}$, and Bart~Sor\'ee$^{3,}$$^{4,}$$^{5,}$$^a$}
\address{$^1$ Department of Materials Science and Engineering, The University of Texas at Dallas, 800 W Campbell Rd., Richardson, Texas 75080, USA.}
\address{$^2$Department of Materials Engineering, KU Leuven, Kasteelpark Arenberg 44, 3001 Leuven, Belgium.}%
\address{$^3$ Imec, Kapeldreef 75, 3001 Heverlee, Belgium.}%
\address{$^4$Department of Electrical Engineering, KU Leuven, Kasteelpark Arenberg 10, 3001 Leuven, Belgium.}%
\address{$^5$Department of Physics, Universiteit Antwerpen, Groenenborgerlaan 171, 2020 Antwerp, Belgium.}
\address{$\rm ^a$Email: bart.soree@imec.be}

\begin{abstract}


We compare three first-principles methods of calculating the Curie temperature in two-dimensional (2D) ferromagnetic materials (FM), modeled using the Heisenberg model, and propose a simple formula for estimating the Curie temperature with high accuracy that works for all common 2D lattice types.
First, we study the effect of exchange anisotropy on the Curie temperature calculated using the Monte-Carlo (MC), the Green's function method, and the renormalized spin-wave (RNSW).
We find that the Green's function overestimates the Curie temperature in high-anisotropy regimes compared to MC, whereas RNSW underestimates the Curie temperature compared to the MC and the Green's function.
Next, we propose a closed-form formula for calculating the Curie temperature of 2D FMs, which provides an estimate of the Curie temperature greatly improving over the mean-field expression for magnetic material screening.
We apply the closed-form formula to predict the Curie temperature 2D magnets screened from the C2DB database and discover several high Curie temperature FMs with $\rm Fe_2F_2$ and $\rm MoI_2$ emerging as the most promising 2D ferromagnets.
Finally, comparing to experimental results for \CrI, \CrCl, and \CrBr,  we conclude that for small effective anisotropies, the Green's function-based equations are preferable, while, for larger anisotropies MC-based results are more predictive.

\end{abstract}
\maketitle
\section{Introduction}
Thanks to the recent discovery of two-dimensional (2D) magnets \CrI~\cite{CrI_exp}, \CrBr~\cite{CrBr_exp}, and \CrGeTe~\cite{CrGeTe_exp}, research in the field of 2D magnets has garnered unprecedented attention in the past few years.
Their percieved application in spintronics\cite{sample21,Nat_comm}, valleytronics~\cite{sample37}, and skyrmion~\cite{Nat_comm2}-based magnetic memories~\cite{sample22} has sparked great interest.
Moreover, the experimental demonstration of the electric field control of the magnetic order in~\CrI~\cite{CrI_exp} provides a path towards the technological realization of electrically tunable magnetic memories using 2D magnets.

However, the low Curie temperature of 2D magnets acts as a hurdle in their practical application.
Most of the 2D magnets discovered experimentally, have a low Curie temperature, \eg, 45 K for \CrI~\cite{CrI_exp} and 34 K for \CrBr~\cite{CrBr_exp}.
While, $\rm Fe_3GeTe_2$ has a Curie temperature of 130 K~\cite{Fe3GeTe2}, it has an itinerant magnetic behavior, which cannot be controlled using an external electric field.
On the other hand, the high Curie temperature in $\rm VSe_2$~\cite{VSe2_good} is a matter of debate with reports emerging of $\rm VSe_2$ having a charge-density-wave ground state with no magnetic ordering~\cite{VSe2_bad,VSe2_bad2}.

The dearth of high Curie temperature 2D magnets has led to an unprecedented effort in the search for 2D magnets with higher Curie temperature.
Thankfully, the possible span of 2D magnets is quite large starting from 2D crystals~\cite{high_throughput_1} to conventional 2D materials doped with transition metals~\cite{Reyntjens_2020,TMD_1,TMD_2,Tiwari2021,PdSeTe,Reyntjens2021}.
However, experiments can only be performed for the most promising 2D ferromagnets.
Hence, a vast amount of research is dedicated to high-throughput screening of the most promising 2D magnets  from theory~\cite{Torelli_2020,Torelli_2019,Kabiraj_2020}.

A common strategy in such high-throughput calculations for predicting the Curie temperature of 2D magnets is: First, obtain 2D materials with magnetic ordering from material databases such as the C2DB~\cite{C2DB}. 
Then, approximate the magnetic structure using a parameterized Heisenberg model whose parameters are obtained from the density-functional-theory (DFT) calculations~\cite{my-paper}.
Finally, predict the Curie temperature from the phase change of the Heisenberg Hamiltonian, calculated using computationally costly Monte-Carlo simulations with anisotropy~\cite{Torelli_2020,Torelli_2019, Kabiraj_2020,citeme2}, or even using the Ising model~\cite{high_throughput_1}.
\new{The Monte-Carlo simulations are costly in terms of computational time and memory compared to the mean-field theory. 
As a result, many researchers use the less accurate mean-field calculations for predicting the Curie temperature of newly discovered materials~\cite{CrI_bad, CrI_bad2}.}

Monte-Carlo simulations with anisotropy result in a rather accurate estimation of the Curie temperature for most of the experimentally verified 2D magnets yet discovered~\cite{my-paper, Torelli_2018}.
However, care must be taken because the calculated Curie temperature depends on the parameters of the Heisenberg Hamitonian~\cite{S_problem},  and on the approximation used at the DFT level.
Moreover, the recent application of methods that take into account the quantum mechanical fluctuations in the Heisenberg model to 2D magnets, \eg, the Green's function method~\cite{Vanherck18,Vanherck20}, and the renormalized spin-wave~\cite{Lado_2017}, raise further questions on how much the Curie temperature depends on the level of approximation used to solve the Heisenberg Hamiltonian.
\new{Moreover, the Curie temperature of 2D ferromagnets depends strongly on the anisotropy~\cite{Mermin66}, which is itself dependent on the spin-orbit interaction of the material \cite{SOC1, SOC2, SOC3,Lado_2017}.
The different methods used to solve the Heisenberg Hamiltonian have a different impact of anisotropy. 
So, it is highly desirable to understand how much the Curie temperature of the 2D ferromagnets depends on various methods used for solving the Heisenberg Hamiltonian.}
\new{There have been previous works on understanding the impact of exchange anisotropy on the Curie temperature of 2D magnets. 
Most of the works have either focussed on using only the MC simulations to obtain a closed form description \cite{citeme2} or have compared the Curie temperature using different methods in extremely high regimes of the exchange anisotropy \cite{Torelli_2018}}.

\fix{We compare, for 34 2D materials, three methods of calculating the Curie temperature from a Heisenberg Hamiltonian: Monte-Carlo (MC), Green's functions, and renormalized spin-waves (RNSW).
We first provide a brief overview of the Heisenberg Hamiltonian, while the three solution methods are discussed elaborately in the appendix.
Next, we provide an analytical formula to approximate the Curie temperature calculated using the three solution methods, as a function of nearest-neighbor exchange strength ($J$) and anisotropy (\DeltaNN).
Further, we calculate the Curie temperature using the three methods, as a function of exchange anisotropy, and fit the analytical formula to each.
We then calculate the Curie temperature of 34 2D ferromagnets screened from the C2DB database~\cite{C2DB} using our analytical formula and find some very promising ferromagnets with high Curie temperature including $\rm Fe_2F_2$, and $\rm MoI_2$ having for all methods estimations above 403 K and 281 K, respectively.
Finally, we show that the Curie temperature calculated using the three methods depends quantitatively on the long-range interactions however, the qualitative trend remains the same, and the analytical formula we developed provides a good estimation for a first theoretical screening.
}

\section{Methodology}\label{s:method}
\subsection{The Heisenberg Hamiltonian}
2D ferromagnets are most commonly modeled through the Heisenberg Hamiltonian
\begin{equation}
	H
	=
	\Over{2}\sum_{i,j}\vec{\hat{S}}_i \tensor{J}_{ij} \vec{\hat{S}}_j
	+ \sum_{i}D (\hat{S}_i^z)^2,
\end{equation}
where, $\vec{\hat{S}}=\hat{S}^x\vec{x}+\hat{S}^y\vec{y}+\hat{S}^z\vec{z}$ is the spin-operator.
Here, the spin-operator can take eigenvalues $S = n/2$, with $n$ a strictly positive integer.
The off-diagonal elements of the J tensor $\tensor{J}_{ij}$ between spins at site $i$ and $j$ has been found to be much smaller than the diagonal elements for most of the 2D FMs~\cite{Xu2018}, hence, we assume the off-diagonal elements to be zero when modelling ferromagnets.
For the present study, exchange interactions up to second neighbor are accounted for, while the second term of the Heisenberg Hamiltonian---called the onsite-anisotropy---is ignored.
The Heisenberg Hamiltonian thus reduces to
\begin{subequations}
\begin{align}
H
&=
	\Over{2}\sum_{i,j}\hat{S}^x_i J^{xx}_{ij} \hat{S}^x_j+\hat{S}^y_i J^{yy}_{ij} \hat{S}^y_j+\hat{S}^z_i J^{zz}_{ij} \hat{S}^z_j\\
&=
	\sum_{i,j} \frac{J_{ij}}{2} \left[%
		\vec{\hat{S}}_i \bcdot \vec{\hat{S}}_j + \Delta_{ij} (\hat{S}^z_i \hat{S}^z_j - \hat{S}^x_i \hat{S}^x_j - \hat{S}^y_i \hat{S}^y_j)
	\right],
	\label{e:Heisenberg}
\end{align}
\end{subequations}
where the anisotropy is modeled through distinct values for the in-plane and out-of-plane anisotropic exchange strength, respectively $J^{xx}=J^{yy}=J(1-{\Delta})$ and $J^{zz}=J(1+{\Delta})$.
The unitless anisotropy $\Delta_{ij}=\infrac{(J^{zz}_{ij}-J^{xx}_{ij})}{2 J_{ij}}$ is said to be of the easy-axis type when positive, while of the easy-plane type otherwise. 

We solve the Heisenberg Hamiltonian as a function of temperature using, the MC, the Green's function, and the RNSW.
The three methods used to solve the Heisenberg Hamiltonian are discussed extensively in the appendix~\ref{s:method}.

\begin{figure}[th]
	\centering   
   \includegraphics[width=\columnwidth]{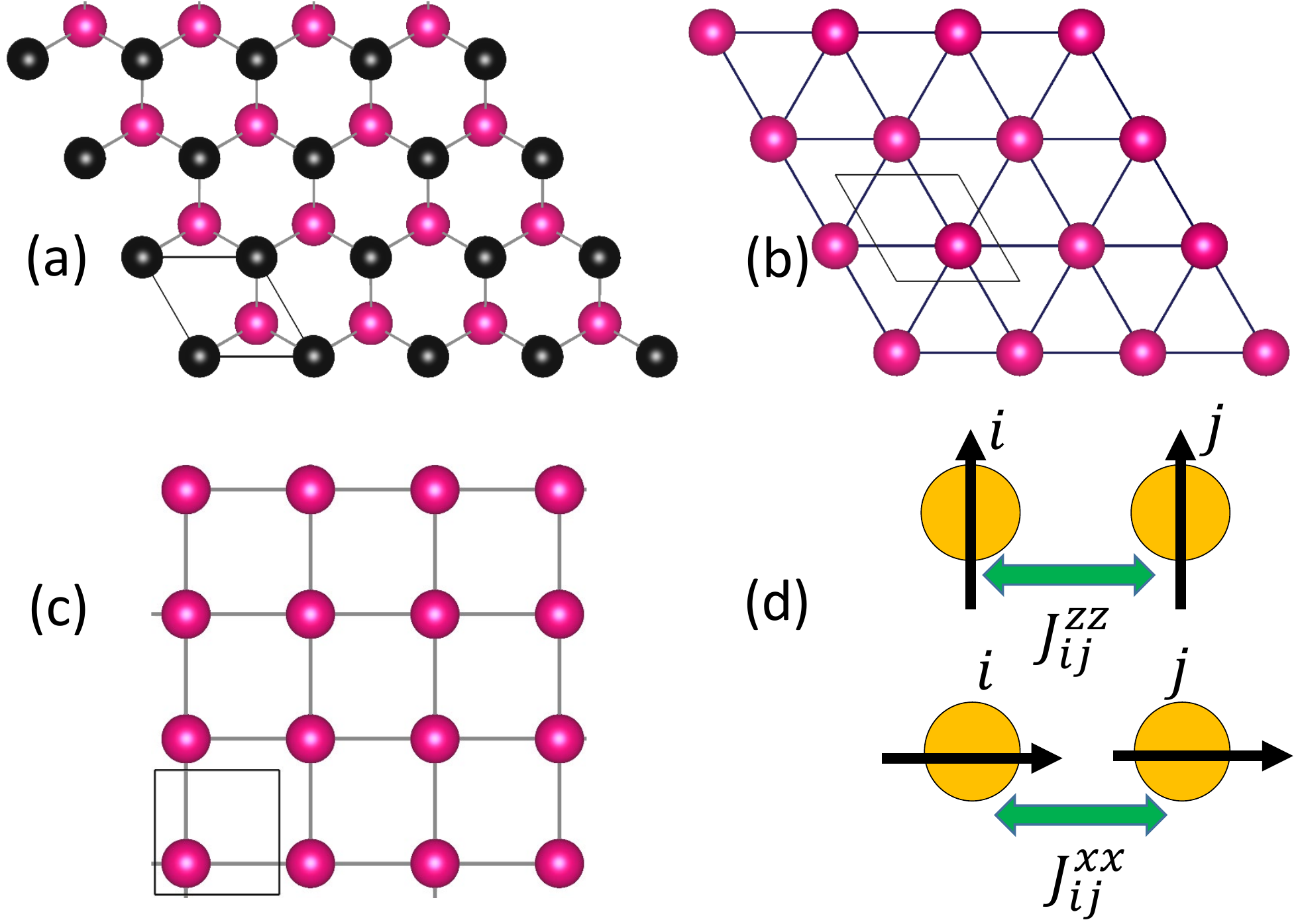} 
    \caption{Three different type of lattices for 2D materials (a) honeycomb, (b) hexagonal, and (c) square lattice. For the hexagonal (b) and the square (c) lattice, the unit cell contains only one atom, whereas for the honeycomb lattice (a), the unitcell has two atoms, referred to as sublattice A (colored pink) and sublattice B (colored black). (d) Visually depicts the in-plane ($J^{xx}_{ij}$) and the out-of-plane ($J^{zz}_{ij}$) exchange interaction between spins $i$ and $j$.}
	\label{f:lattices}
\end{figure} 
\subsection{Analytical formula for screening 2D magnets}
The exact methods of calculating Curie temeprature using the MC, the Green's function and the RNSW are computationally costly.
Hence, to provide a closed form equation for \TCurie~calculated using all three methods, we propose the analytical formula, 
\begin{equation}
T_{\mathrm{C}}=\frac{1}{\alpha_1-\alpha_2\ln(\Delta_{\mathrm{NN}})}\frac{J(S^2+\theta S)}{k_{\rm B}},
\label{e:empirical}
\end{equation}
which is inspired on the group theoretical approach used by Bander et.al~\cite{Bander88}.
Here, $S$ is the spin eigen value and $\Delta_{\rm NN}$ is the nearest-neighbor exchange anisotropy. 
$\theta=1$ for Green's function, and RNSW, and $\theta=0$ for MC.
The MC results in a Curie temperature scaling by $S^2$, whereas for RNSW and Green’s function, the Curie temperature scales at $S(S+1)$. 
Therefore, for larger $S$, the MC will result in a lower Curie temperature than Green’s function and RNSW.
Dimensionless parameters $\alpha_1$ and $\alpha_2$ are fit so that \TCurie~matches the Curie temperature as a function of $\Delta_{\mathrm{NN}},\,J$, and $S$, obtained using MC, Green's, and RNSW for the hexagonal, honeycomb, and square lattice (Fig.~\ref{f:lattices}).

\new{The anisotropy originates due to the spin-orbit coupling, and spin-orbit coupling is a much weaker interaction than the electronic exchange interaction. 
$\Delta_{\rm NN}$ is the ratio of the strength of anisotropy and exchange interaction $J$, it is highly unlikely that $\Delta_{\rm NN}$ would approach 1 for any 2D magnetic material.
Nevertheless, we fit the formula only up to $\Delta_{\rm NN}=0.2$ to the exact methods.  
Also, the group-theoretical method applied by Bander et.al~\cite{Bander88} is only valid around $\Delta_{\rm NN} \rightarrow 0$ from which our formula is inspired.
Therefore, the formula should only be used to calculate the Curie temperature of materials whose $\Delta_{\rm NN}\leq0.2$.}

\subsection{Obtaining the input parameters of the analytical formula}
The parameters for the analytical formula proposed in Equation~(\ref{e:empirical}) can be obtained from DFT total energy calculations using,
\begin{subequations}
	\begin{align}
		J^{\perp} &=\frac{E^{\perp}_{\mathrm{FM}}-E^{\perp}_{\mathrm{AFM}}}{2N_{\mathrm{NN}}S^2},\\
		J^{\parallel} &=\frac{E^{\parallel}_{\mathrm{FM}}-E^{\parallel}_{\mathrm{AFM}}}{2N_{\mathrm{NN}}S^2},\\
        J &=\frac{J^{\perp}+J^{\parallel}}{2},\\
        \Delta_{\mathrm{NN}} &=\frac{J^{\perp}-J^{\parallel}}{2J}.
    \label{e:J}
\end{align}
\end{subequations}
Here, $E^{\perp/\parallel}_{\mathrm{FM}}$ and $E^{\perp/\parallel}_{\mathrm{AFM}}$ are the total energies calculated using the DFT for FM and AFM order with magnetic axis oriented in the out-of-plane/in-plane direction.
$N_{\mathrm{NN}}$ is the number of nearest neighbors.
For obtaining parameters beyond nearest-neighbor, one has to use advanced mapping methods presented in Ref.~\onlinecite{my-paper}.

\begin{figure*}[th]
	\centering   
   \subfigure[]{\includegraphics[width=\columnwidth]{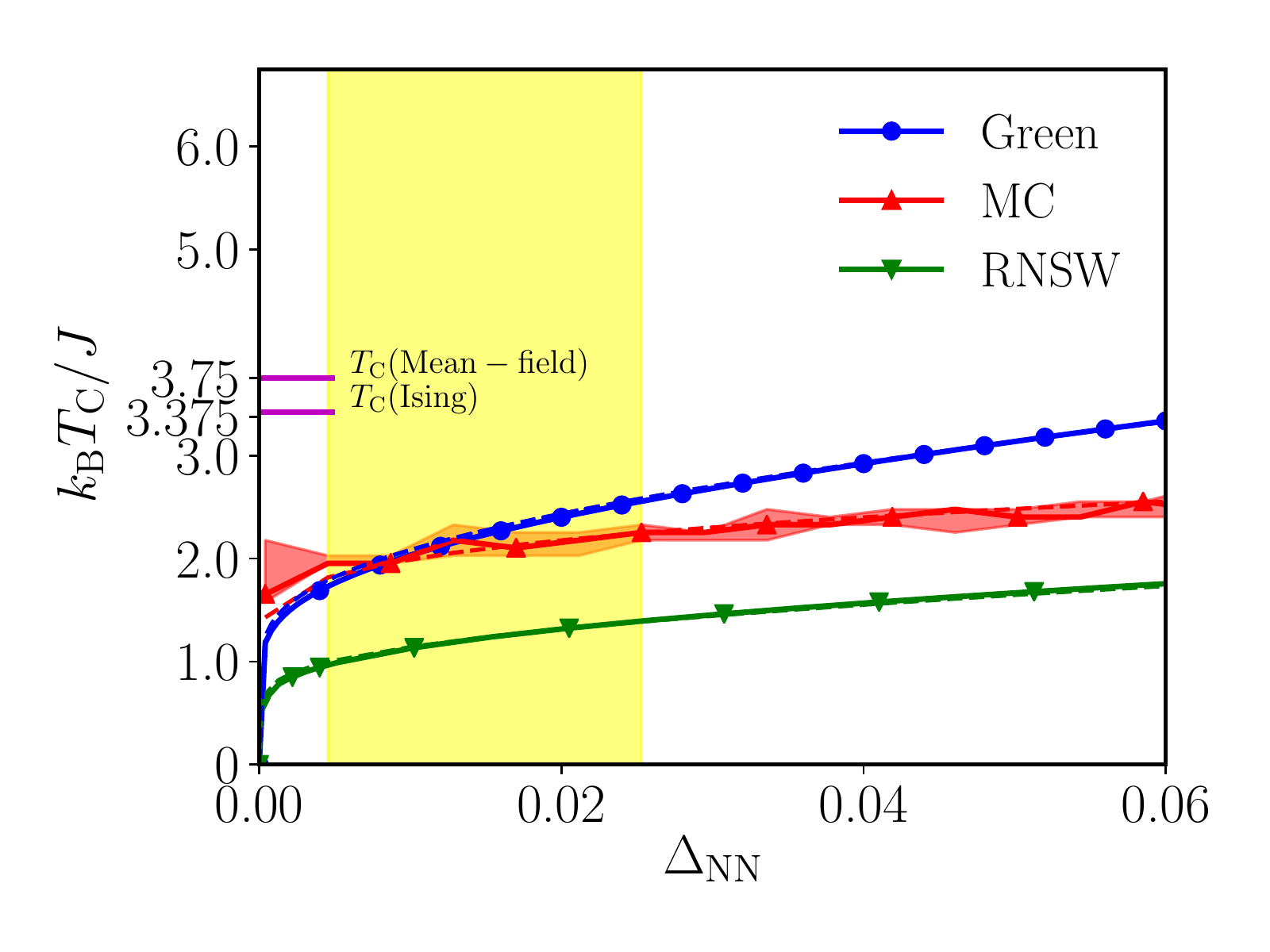}} 
   \subfigure[]{\includegraphics[width=\columnwidth]{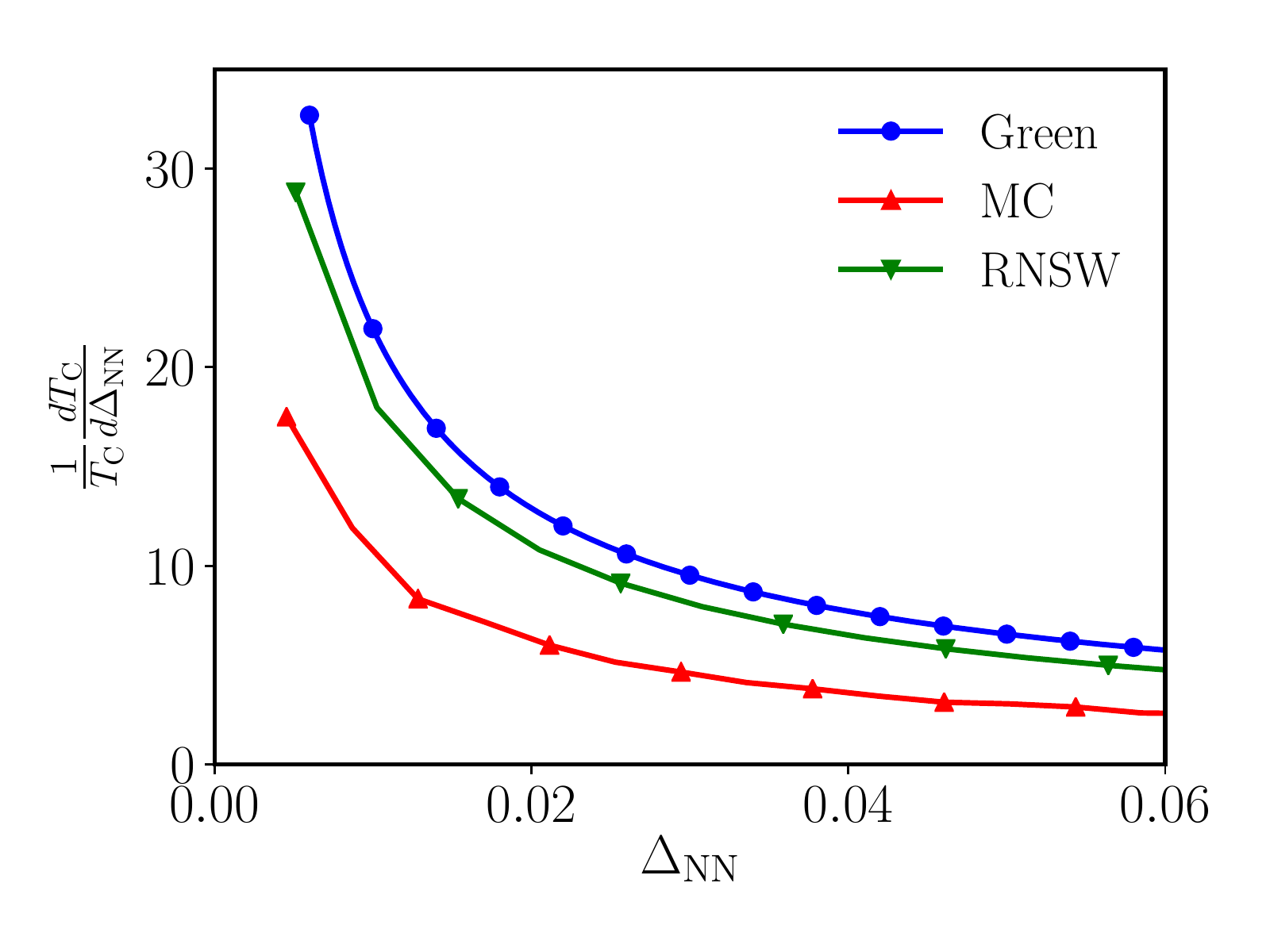}}
    \caption{(a) Comparison of Curie temperature calculated using the Green's function, the Monte-Carlo (MC), and the renormalized spin-wave (RNSW) method as a function of nearest-neighbor exchange anisotropy ($\Delta_{\mathrm{NN}}$) for a honeycomb lattice. \new{For the MC, the solid line shows the median and the shading shows the 25th to 75th percentile of the calculated Curie temperature.The dashed lines show a fit using function $(\alpha_1-\alpha_2\ln(\Delta_{\mathrm{NN}}))^{-1}J(S^2+\theta S)/k_{\rm B}$ for the respective methods. The horizontal tick shows the Ising limit ($1.52 S^2$, with $S=3/2$) and the quantum mean-field ($1/3(S(S+1)N_{\rm NN})$, $N_{\rm NN}$=3 for honeycomb lattice).} The yellow shaded area shows the region where the MC and the Green's function have a difference less than 10 \%. (b) Comparison of Curie temperature sensitivity calculated using the Green's function, the MC, and the RNSW method as a function of nearest-neighbor exchange anisotropy ($\Delta_{\mathrm{NNN}}$) for a honeycomb lattice.}
	\label{f:Curie_NN}
\end{figure*} 
\begin{figure*}[t]
  \centering  
  \subfigure[]{\includegraphics[width=2\columnwidth]{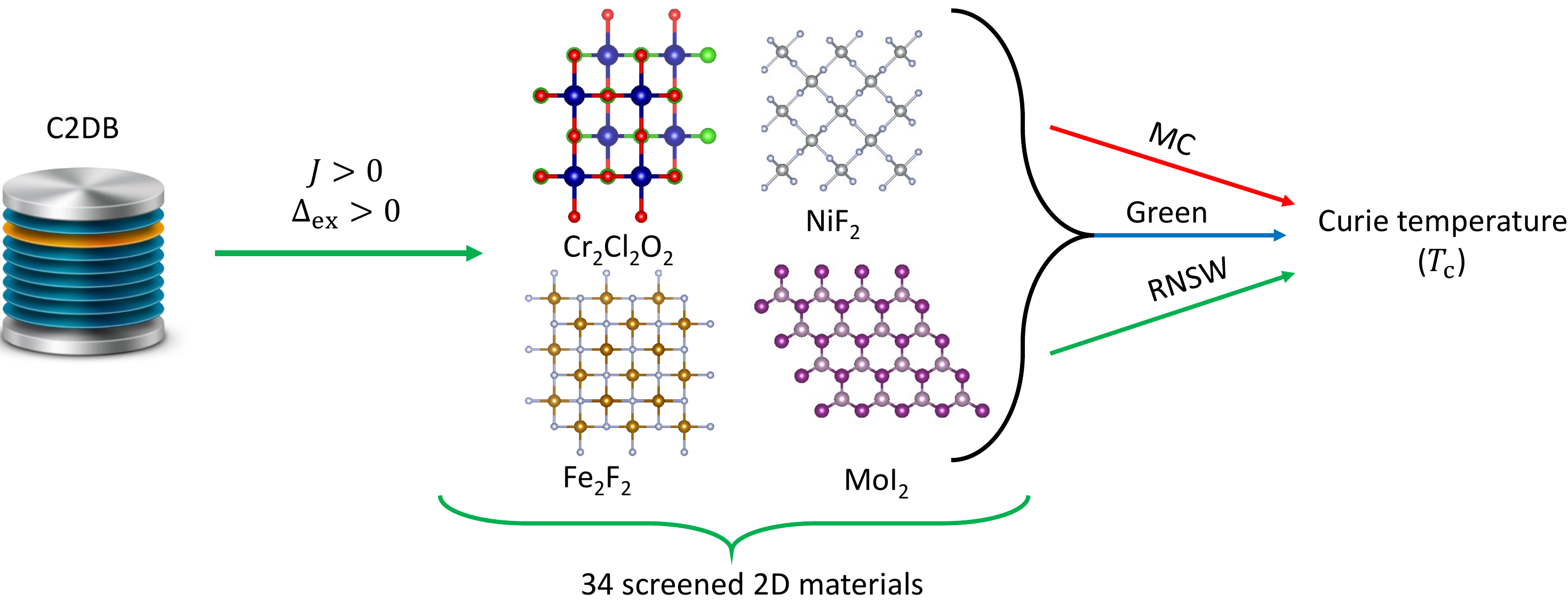}}
  \subfigure[]{\includegraphics[width=1\columnwidth]{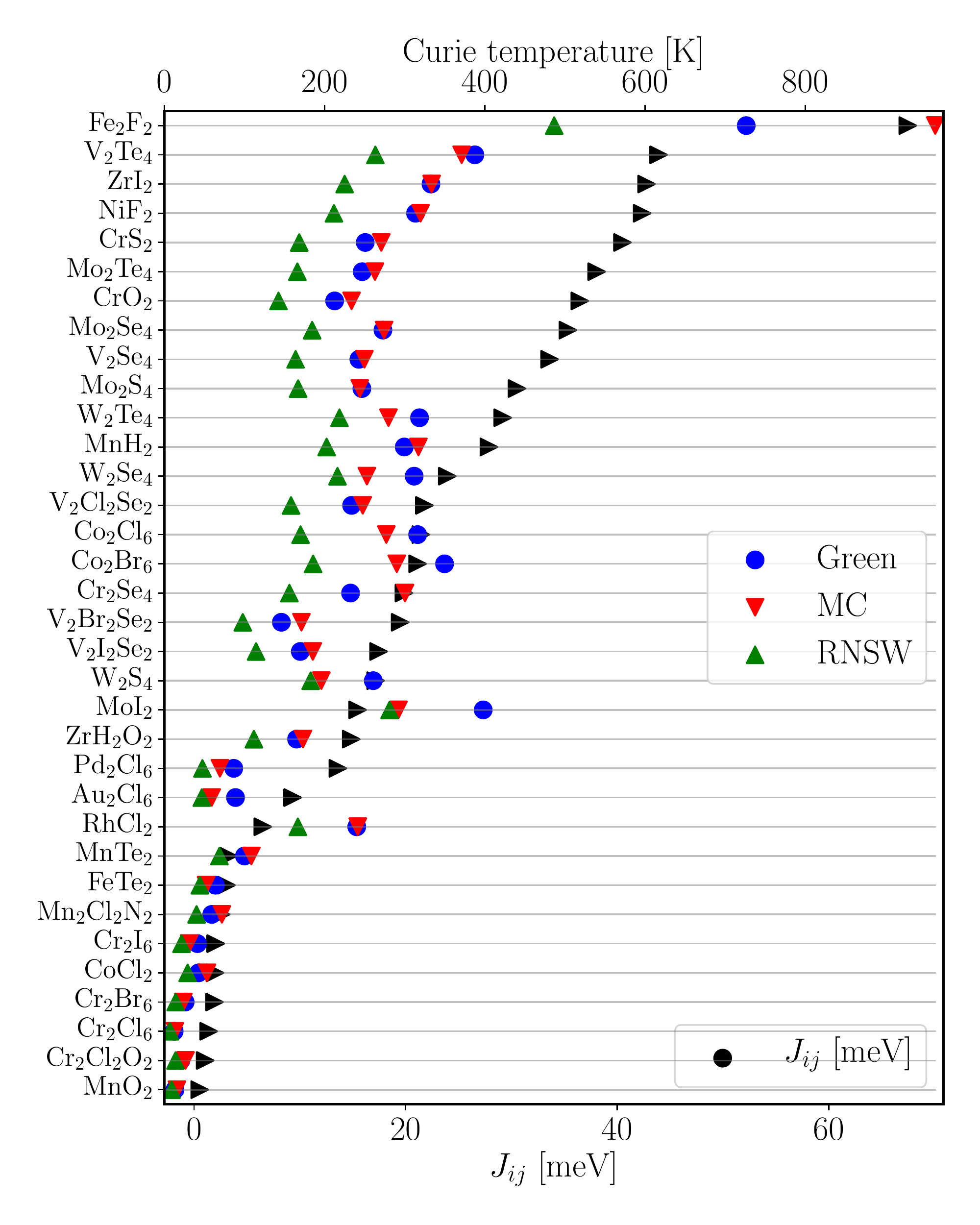}}
  \subfigure[]{\includegraphics[width=1\columnwidth]{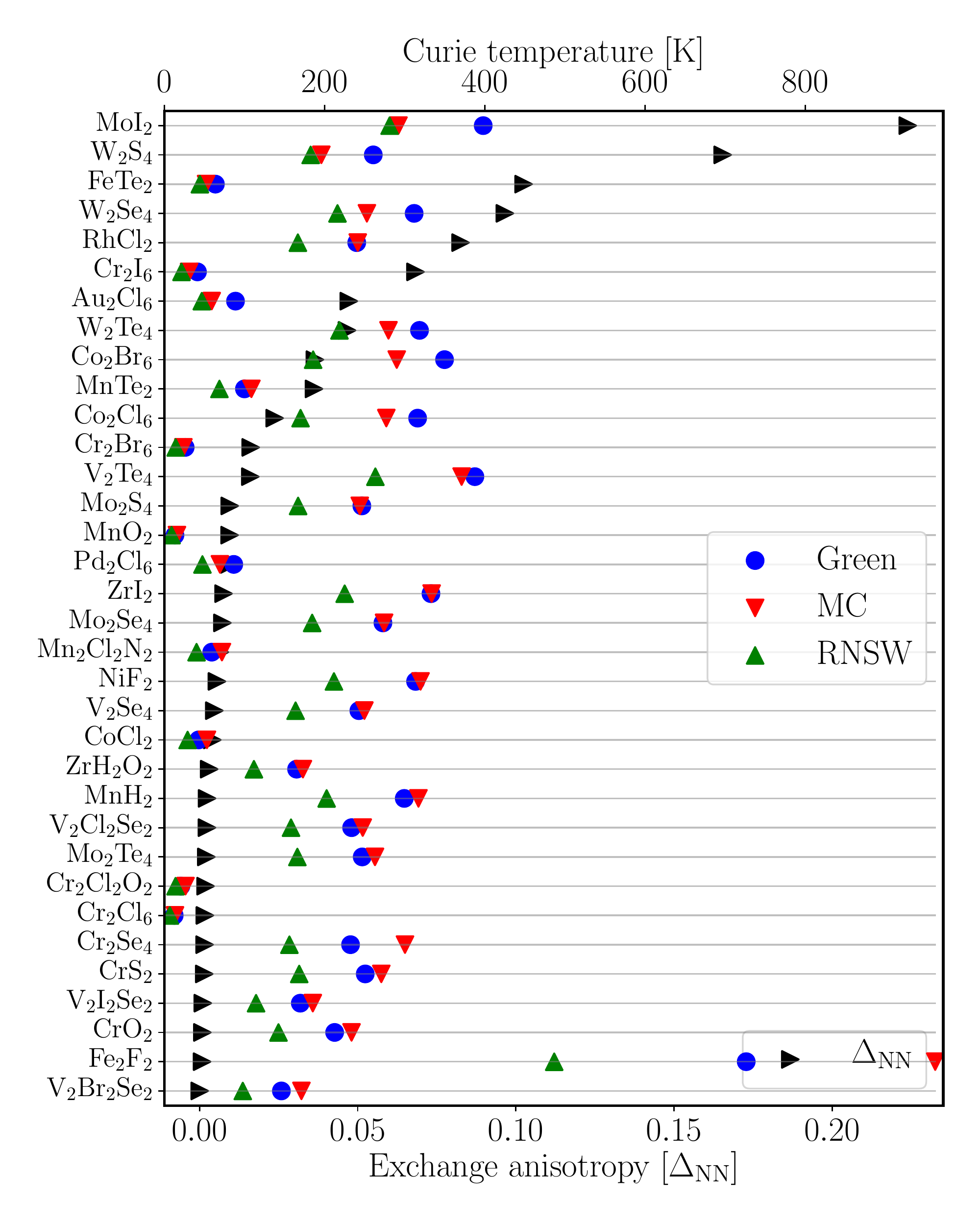}}
  \caption{(a) Materials obtained from the C2DB database with positive out-of-plane anisotropy and positive exchange interaction. The Curie temperature of the screened materials using the analytical formula for MC, the Green's function, and the RNSW (top abscissa) sorted as a function of (a) exchange interaction ($J_{ij}$), (b) exchange anisotropy ($\Delta_{\rm NN}$).} 
  \label{f:Curie_screen} 
\end{figure*}

\section{Results and Discussion}\label{s:results}

We first discuss the impact of nearest-neighbor exchange anisotropy on the Curie temperature calculated using the three exact methods, and fit to this the \TCurie~calculated using the analytical formula. 
We then calculate the Curie temperature of 2D ferromagnets screened from C2DB database using the fitted analytical formula.
Finally, we discuss the impact of next-nearest neighbor anisotropy on the Curie temperature of Cr-compounds.

\subsection{Effect of exchange anisotropy}\label{s:exchange}

Figure~\ref{f:Curie_NN} (a) shows the calculated Curie temperature (\TCurie) for a hexagonal 2D material as a function of nearest-neighbor exchange anisotropy ($\Delta_{\mathrm{NN}}$) using the three methods.
For all three methods, the Curie temperature increases with increasing anisotropy.
Moreover, for zero anisotropy, the Curie temperature tends to zero for all the three methods.

There are three regions with a distinct feature in Fig.~\ref{f:Curie_NN} (a).
First, within the yellow shaded region, the Curie temperature calculated using the MC and the Green's function matches closely.
\new{For lower anisotropies, the Green's function approach results in a Curie temperature lower compared to the MC, whereas for higher anisotropies, the situation is opposite: the Curie temperature estimated using the Green's function is higher than the one estimated using the MC.}
Moreover, for all three regions, renormalized spin-wave results in a Curie temperature below both the MC and the Green's function.

The existence of three regions for the MC and the Green's function can be understood by careful observation of the exact Curie temperature formulas.
In the limit of $\Delta_{\mathrm{NN}}\rightarrow1$, the Curie temperature calculated using the Green's function tends to the result from the molecular-field theory~\cite{Vanherck18}, which overestimates the Curie temperature.
Whereas for higher anisotropies $\Delta_{\mathrm{NN}}\rightarrow1$, the Curie temperature calculated using the MC remains below the Ising limit~\cite{approach_1}. 
\new{Therefore, because the molecular field theory results in an overestimation of the Curie temperature compared to the Ising model, the Green's function overestimates the Curie temperature compared to MC at higher anisotropies.}
For $\Delta_{\mathrm{NN}}=0$, the Curie temperature calculated using the Green's function goes to zero in accordance the Mermin-Wagner theorem~\cite{sample27}, whereas for MC, the average magnetization $M=\sqrt{S_x^2+S_y^2+S_z^2}$ remains finite due to non-zero exchange interaction.
Hence, for lower anisotropies, MC overestimates the Curie temperature. 

To further understand the impact of $\Delta_{\mrm{NN}}$, we plot the sensitivity ($\frac{1}{T_{\mrm{C}}}\frac{dT_{\mrm{C}}}{d\Delta_{\mrm{NN}}}$) of the Curie temperature in Fig.~\ref{f:Curie_NN} (b) for all three methods. 
We observe that both the RNSW and Green's function have almost similar sensitivity to $\Delta_{\mrm{NN}}$, and the sensitivity decreases with increasing anisotropy. 
Moreover, we see that MC is much less sensitive to $\Delta_{\mrm{NN}}$, especially at lower anisotropy compared to RNSW and Green's function.

\subsection{Fitting of analytical formula to exact methods}
\begin{table}[ht]
\caption{Parameters of the analytical formula} 
\centering
\resizebox{0.75\columnwidth}{!}{
\begin{tabular}{c  c  c  c c} 
\toprule
Lattice & Parameter & MC & Green & RNSW \\ [0.1ex]
\midrule
\multirow{2}{*}{Honeycomb}  & $\alpha_1$ & 0.49 & 0.07 & 0.40    \\
 & $\alpha_2$ & 0.14 & 0.37 & 0.62   \\
\\
\multirow{2}{*}{Hexagonal} &$\alpha_1$ & 0.24 & 0.24 & 0.32    \\
  & $\alpha_2$ & 0.045 & 0.14 & 0.21   \\
\\
\multirow{2}{*}{Square} & $\alpha_1$ & 0.37 & 0.34 & 0.43     \\
  & $\alpha_2$ & 0.08 & 0.24 & 0.36 \\
\bottomrule
\end{tabular}
}
\label{t:log_formula}
\end{table}

Figure~\ref{f:Curie_NN} also shows the fit of the analytical function (Eq.~(\ref{e:empirical})) to the Curie temperature calculated using the MC, the Green's function, and the RNSW, respectively for a honeycomb lattice. 
Corresponding figures for the hexagonal lattice and the square lattice are provided in supplementary document (supplemetary.pdf). 
For all three lattices, we use $\Delta_{\mathrm{NN}}=0.0$ to $\Delta_{\mathrm{NN}}=0.2$ as the fitting range.
We observe that Eq.~(\ref{e:empirical}) fits remarkably well to the Curie temperature calculated using the mentioned methods.
The parameters $\alpha_1$ and $\alpha_2$ are provided in Table~\ref{t:log_formula}.
From Table~\ref{t:log_formula}, we see that the parameter $\alpha_2$ for MC is much lower compared to Green's function and RNSW.

\subsection{Screening of 2D magnets from C2DB and their critical temperatures}

Figure~\ref{f:Curie_screen} (a) shows a schematic of our screening process.
We screen the C2DB database~\cite{C2DB} for ferromagnetic 2D materials with out-of-plane exchange anisotropy ($J>0$ and $\Delta_{\mathrm{NN}}>0$).
We find 34 2D magnets with FM order in their ground state.
Next, we calculate their Curie temperature using the analytical formula in Eq.~(\ref{e:empirical}).

Figure~\ref{f:Curie_screen} (b) and (c) show the calculated Curie temperature of the screened ferromagnets sorted as a function of exchange interaction ($J_{ij}$) and exchange anisotropy ($\Delta_{\mathrm{NN}}$), calculated using the MC, the Green's function, and the RNSW.
The table comprising the Curie temperature using different methods is provided in the supplementary document (supplementary.pdf).

We observe some general trends from Fig.~\ref{f:Curie_screen} (b) and (c).
Firstly, the Curie temperature is indeed dependent on the methods used to solve the Heisenberg Hamiltonian.
We observe the same pattern as in Fig.~\ref{f:Curie_NN} (a), where the Curie temperature calculated using RNSW remains low for all the materials, however, for higher anisotropy, RNSW starts approaching the MC.
Given that all the identified 2D ferromagnets from C2DB database have anisotropy: $\Delta_{\mathrm{NN}}<0.25$, we can say that RNSW gives the most conservative estimation of the Curie temperature.
On the other hand, the Curie temperature calculated using the Green's function remains below the MC for compounds till $\rm ZrI_2$ ($\Delta_{\mathrm{NN}}=0.0072$).
Most remarkably, from our screening using the analytical formulas for all three methods, we identify some very-promising candidates for realizing 2D ferromagnets are $\rm Fe_2F_2$ and $\rm MoI_2$ for whom even RNSW predicts a Curie temperature 403 K and 231 K, respectively.
$\rm Fe_2F_2$ has a high exchange interaction strength that leads to its higher Curie temperature, whereas $\rm MoI_2$ has a higher exchange anisotropy that leads to a higher Curie temperature.

Interestingly, we see from Fig.~\ref{f:Curie_screen} (b) and (c) that all three experimentally discovered Cr-compounds: \CrI, \CrBr, and \CrCl~are screened from the C2DB database.
For \CrI, we obtain a Curie temperature of 41 K, 31 K, and 23 K (45 K experimental~\cite{CrI_exp}), and for \CrBr, we find a Curie temperature of 26 K, 24 K, and 14 K (34 K experimental~\cite{CrBr_exp}) from the Green's, the MC, and the RNSW method, respectively.
The close estimation of the Curie temperature of Cr-compounds compared to their experimental values, suggests that the analytical formula in Eq.~(\ref{e:empirical}) can be used for an efficient first level screening of 2D ferromagnetic compounds.

\new{It should be noted that the obtained high Curie temperature for metals e.g., $\rm Fe_2F_2$ using our formulas provides a rough first-level estimate because their itinerant magnetic nature is not fully captured by the Heisenberg model. 
Therefore, a more detailed approach is needed to precisely predict their magnetic order as a function of temperature e.g., dynamical mean-field theory}


\subsection{Impact of next-nearest neighbor anisotropy}
Until now, we discussed the impact of nearest-neighbor exchange anisotropy on the Curie temperature calculated using various methods and showed our analytical formula, used for high-throughput screening of 2D FMs.
\new{Here, we discuss the impact of next-neighbor interactions, especially the next-neighbor anisotropy, and the impact of long-range interactions on the theoretical prediction of Curie temperature of 2D FMs. 
We use the honeycomb lattice as an example.}

\begin{figure}[t]
	\centering
   \includegraphics[width=\columnwidth]{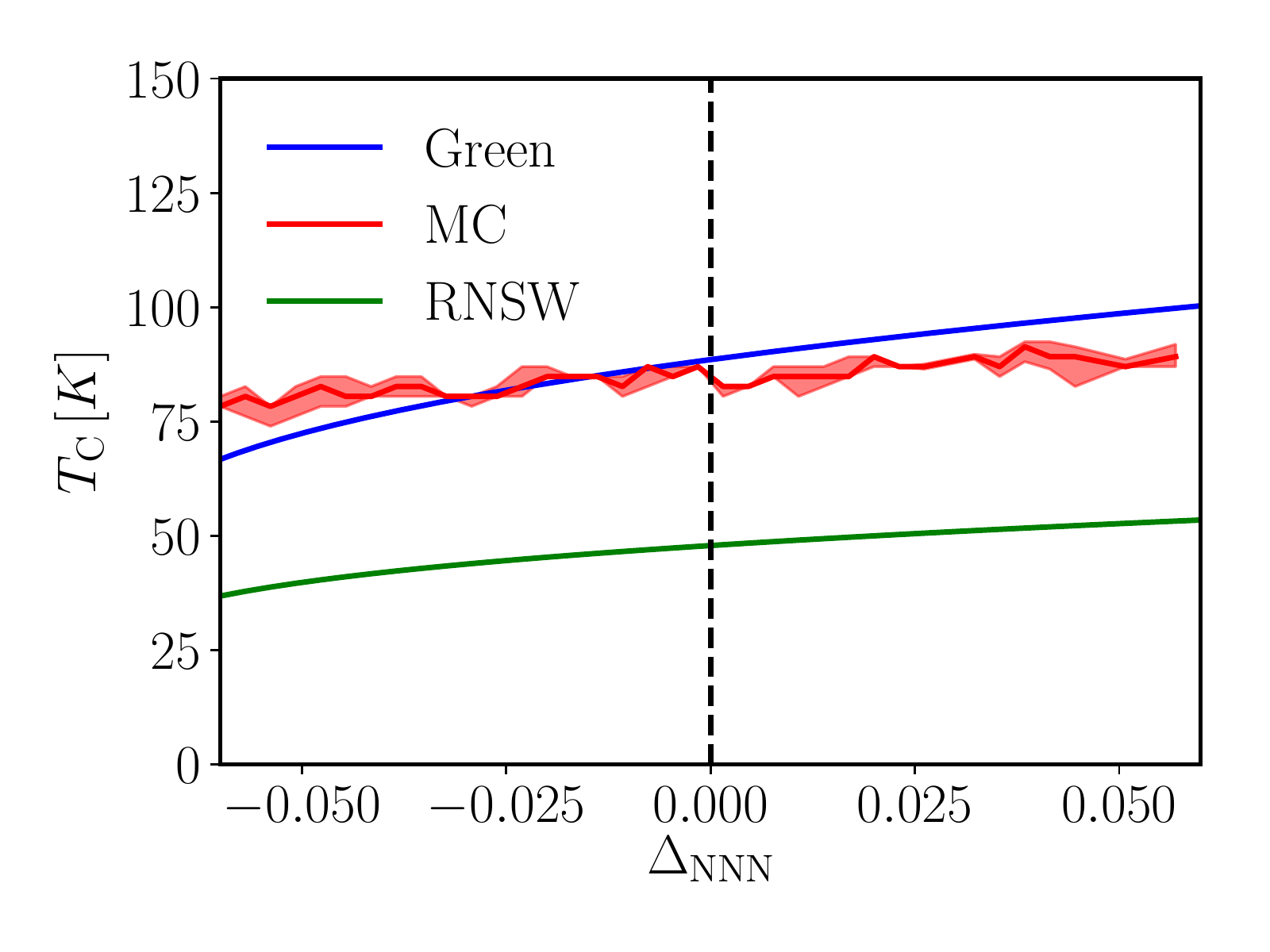}
    \caption{Comparison of Curie temperature calculated using Green's function, Monte-Carlo and the renormalized spin-wave method as a function of next-nearest-neighbor exchange anisotropy ($\Delta_{\mathrm{NNN}}$) for a honeycomb lattice. The vertical dashed line shows the $\Delta_{\mathrm{NNN}}=0$ at which Fig.~\ref{f:Curie_NN} is calculated. \new{For the MC, the solid line shows the median and the shading shows the 25th to 75th percentile of the calculated Curie temperature.}}
	\label{f:Curie_NNN}
\end{figure}

Figure~\ref{f:Curie_NNN} shows the comparison between the Curie temperature (\TCurie) for a honeycomb 2D material as a function of next-nearest-neighbor exchange anisotropy ($\Delta_{\mathrm{NNN}}=\infrac{(J_{\mrm{NNN}}^z-J_{\mrm{NNN}}^x)}{(2J_{\mrm{NNN}})}$) for a honeycomb lattice. 
We keep the nearest-neighbor anisotropy fixed at $\Delta_{\mrm{NN}}=0.01$, and nearest-neighbor exchange at $J_{\mrm{NN}}=2.5$ meV.
We observe that both the Green's function and RNSW have a similar sensitivity to the next-nearest-neighbor anisotropy, whereas MC calculations are relatively less sensitive.
As dicussed in the section~\ref{s:exchange}, the observed low sensitivity of MC simulations to next-neighbor anisotropy is due to their classical nature.
\subsection{Importance of long-range interactions for the quantitative estimation of \TCurie}
While the approximate results of the analytical formula are good as a first screening, further neighbours can still lead to significant changes of the \TCurie.
We compare the Curie temperature calculated for Cr-compounds: \CrI, \CrBr, \CrCl, and \CrGeTe, using the MC, the Ising MC, the Green's function, and the RNSW, to their experimentally measured value in Table II, including the long-range interactions. 
The experimental comparison of methods is subject to the parameters obtained from DFT and with a change in exchange functional, the J-parameters change. However, the PBE parameters have been found to be close to experiments for CrI3~\cite{neutron-diffraction} and serve as a good benchmark for evaluating the three methods and their regimes of applicability.

\begin{table}[ht]
\caption{Curie Temperature [K] of Cr-compounds using exact methods with long-range interactions} 
\centering
\resizebox{0.9\columnwidth}{!}{
\begin{tabular}{c c c c c c c} 
\toprule
Method  & MC & Green & RNSW & Ising & Mean-Field (NN) & Exp\\ [0.5ex]
\midrule
\CrI & 63 & 83  & 36  & 181  & 89 &  45 ~\cite{CrI_exp}  \\
\CrBr & 37  & 39  & 20   &  130  & 55 & 35 ~\cite{CrBr_exp}  \\
\CrCl &  25  & 21  & 15 &  100  & 53 & 17 (bulk)~\cite{CrCl3_exp_bulk}   \\
\CrGeTe &  64   & 68  & 38   &  314  & 237 & 42  (bi-layer)~\cite{CrGeTe_exp}  \\[1ex]
\bottomrule
\end{tabular}
}
\label{t:Curie_table}
\end{table}

We observe that the Curie temperature calculated using the MC and the RNSW methods is in good agreement with the experimental values for \CrI, and MC and the Green’s function methods is in good agreement for \CrBr.
Whereas the Curie temperature calculated using the RNSW and the Green's function results in a good agreement for \CrCl.
The Curie temperature of monolayer \CrGeTe~has not been reported experimentally yet, however, if we consider that the Curie temperature follows the same trend as reported in~\cite{CrGeTe_exp}, we find that RNSW results in a much closer estimation of the Curie temperature.
It is worth noting that the experimental Curie temperature for all three Cr-compounds lies between what is calculated by the three methods with RNSW setting the lower limit and Green’s function and MC setting the upper limit.

\begin{table}[ht]
\caption{$J$-parameters and anisotropies of experimental Cr-compounds} 
\centering
\resizebox{0.9\columnwidth}{!}{
\begin{tabular}{c c c c c c c} 
\toprule
Parameters  & $J_{\rm NN}$ &$J_{\rm NNN}$ & $J_{\rm NNNN}$& $\Delta_{\rm NN}$ &$\Delta_{\rm NNN}$ & $\Delta_{\rm NNNN}$ \\ [0.5ex]
                   & (meV) & (meV) & (meV) &  & & \\ [0.5ex]
\midrule
\CrI & 2.21 & 0.75  & -  & 0.029  &  0.045 &-  \\
\CrBr & 1.38 & 0.44  & -  & 0.010  &  0.012&-  \\
\CrCl & 1.31 & 0.24  & -  & 0.001 &  0.008 &-  \\
\CrGeTe &5.87 & -0.28  & 0.345 & 0.02 &  0.0 & 0.028  \\[1ex]
\bottomrule
\end{tabular}
}
\label{t:J_table}
\end{table}

We now compare the $J$-parameters for the Cr-compounds shown in Table~\ref{t:J_table} to the Curie temperatures calculated using various methods in Table~\ref{t:Curie_table}.
From Table~\ref{t:J_table}, we find that a nearest-neighbor anisotropy $\Delta_{\rm NN} = 0.067$ for \CrI.
As expected from Fig.~\ref{f:Curie_NN}, the MC Curie temperature is below the Green's function Curie temperature for \CrI.
For \CrBr, the nearest-neighbor anisotropy $\Delta_{\rm NN} = 0.02$, which is in the range where MC and Green's function almost overlap in Fig.~\ref{f:Curie_NN}.
Hence, the Curie temperature calculated using MC is almost the same as one calculated using the Green's function.
For \CrCl, the nearest neighbor anisotropy $\Delta_{\rm NN} = 0.002$, which is in the range where RNSW and Green's function are closer and MC overestimates the Curie temperature.
Hence, the Curie temperature calculated using MC is higher than the Green's function.
For \CrGeTe, the anisotropy is almost the same as \CrBr, hence, the Curie temperature calculated using MC and Green's function is almost the same.

We see that the inclusion of the next-neighbor interaction does impact the Curie temperature quantitatively.
A similar conclusion was drawn in our previous works for both bulk~\cite{my-paper} and monolayer 2D magnets~\cite{Vanherck20}.
However, the trend of Fig.~\ref{f:Curie_screen} remains the same with RNSW being the most conservative and MC/Green's function interchanging their estimation depending on the anisotropy.
Therefore, the analytical formulas provided in this work can be used to obtain a qualitative estimation of Curie temperatures of 2D ferromagnets with the least effort.
However, for quantitatively more accurate results, one will have to include the long-range interactions.

\section{Conclusion}\label{s:conclusion}
We have compared three common methods: the MC, the Green's function, and the RNSW used for calculating the Curie temperature of 2D ferromagnets, which are modeled using the Heisenberg Hamiltonian.
We have investigated the impact of nearest-neighbor exchange anisotropy as well as the next nearest-neighbor anisotropy on the Curie temperature calculated using the mentioned methods.
We have shown that the Curie temperature calculated using the Green's function and the MC method as a function of nearest-neighbor anisotropy result in three regions. 
At low anisotropy, the Green's function results in a lower Curie temperature.
At higher anisotropy, the MC results in a lower Curie temperature, and in-between there is a region where both the Green's function and the MC result in the same Curie temperature estimation.

We have provided a closed-form analytical formula to calculate the Curie temperature of 2D FMs using nearest neighbor exchange and anisotropy.
The analytical formula to calculate the Curie temperature has been fitted to the exact Curie temperature obtained from the MC, the Green's function and the RNSW method.
We have applied our formula on 34 2D ferromagnets screened from the C2DB database~\cite{C2DB} and found some very promising ferromagnets with high Curie temperature including $\rm Fe_2F_2$, and $\rm MoI_2$ having the lowest estimation of 403 K and 281 K, respectively.

\new{By comparing the Curie temperature of the experimentally grown Cr-compounds, we found that for \CrCl, which has low anisotropy, the RNSW and the Green’s function method results in a more accurate estimation of the Curie temperature, whereas for \CrI, which has a higher anisotropy, MC results in a better estimation of the Curie temperature. 
Moreover, for \CrBr, which has an intermediate anisotropy, both the MC and the Green’s function results in a similar estimation of the Curie temperature, suggesting that for materials with low anisotropy, the RNSW and the Green’s function will result in a more accurate description of the Curie temperature, whereas for materials with higher anisotropy, MC results in a better estimation of the Curie temperature.}

Finally, we have shown that the inclusion of long-range interactions does impact the Curie temperature quantitatively, however, the qualitative behavior remains the same.
Therefore, the analytical formulas provided in this work can be used to obtain a qualitative estimation of Curie temperatures of 2D ferromagnets with the least effort.
However, for quantitatively more accurate results, one will have to include the long-range interactions and use the exact methods depending on the anisotropy.

\section{Appendix}{\label{s:method}} 
We briefly discuss the MC, the Green's function, and the RNSW in this section. 
\subsection{Renormalized spin wave}
First proposed by Bloch et.al~\cite{Bloch1930}, spin-waves have been used extensively in the theory of magnetism.
The framework of the spin-waves starts by bosonizing the spin-operators ($\mathbf{\hat{S}}$) of the Heisenberg Hamiltonian using the Holstein-Primakoff transformation~\cite{H-P} with $\hat{S}^{+}=2S\sqrt{1-\frac{\hat{a}^\dagger \hat{a}}{2S}}\hat{a}^\dagger$, $\hat{S}^{-}=2S \hat{a}\sqrt{1-\frac{\hat{a}^\dagger \hat{a}}{2S}}$, and $\hat{S}^{z}=S-{\hat{a}^\dagger a}$.
Here, $\hat{S}^{x}=\frac{\hat{S}^{+}+\hat{S}^{-}}{2}$, and $\hat{S}^{y}=\frac{\hat{S}^{+}-\hat{S}^{-}}{2i}$.
$\hat{a}^\dagger/\hat{a}$ are the bosonic creation/annhilation operators, and $S$ is the spin.
We put the value of spin operators and $J$ in \Eq{e:Heisenberg} and ignore the fourth order terms,
\begin{equation}	
\begin{aligned}
	H=-\sum_{i,j}J_{ij}(1-\Delta_{ij})S(\hat{a}^{\dagger}_i\hat{a}_j-\frac{\hat{a}^{\dagger}_i\hat{a}^{\dagger}_i\hat{a}_i\hat{a}_j}{2S}-\frac{\hat{a}^{\dagger}_i\hat{a}^{\dagger}_j\hat{a}_j\hat{a}_j}{2S})\\
+h.c -\Over{2}\sum_{i,j}J_{ij}(1+\Delta_{ij})(S^2-S\hat{a}^{\dagger}_ia_i+S\hat{a}^{\dagger}_j\hat{a}_j+\hat{a}^{\dagger}_i\hat{a}_i\hat{a}^{\dagger}_j\hat{a}_j).
	\label{e:SW}
\end{aligned}
\end{equation}
We now make a Hartree-Fock approximation to decouple the 2\nth{nd} order terms of \Eq{e:SW} as, ${\hat{a}^{\dagger}_i\hat{a}^{\dagger}_j\hat{a}_j\hat{a}_j}= \hat{a}^{\dagger}_i \hat{a}_j \langle \hat{a}^{\dagger}_j\hat{a}_j\rangle$.
The second order terms are merely the bosonic number density terms ($\langle \hat{a}^{\dagger}_i\hat{a}_i\rangle=\langle n\rangle$).
With these substitutions we obtain,
\begin{equation}	
	H=H_0\left(1-\frac{\langle n \rangle}{2S}\right)
	\label{e:SW2}
\end{equation}
With $H_0=-\sum_{i,j}J_{ij} (1-\Delta) S(\hat{a}^{\dagger}_i\hat{a}_j)+\Over{2}\sum_{i,j}J_{ij}(1+\Delta)(S\hat{a}^{\dagger}_i\hat{a}_i+S\hat{a}^{\dagger}_j\hat{a}_j)+h.c$.
The creation and annhilation operators are transformed in their reciprocal space,
\begin{subequations}
\begin{align}
a^\dagger_i &=\sum_{\mathrm{\mathbf{k}}\in B}\exp(i\mathrm{\mathbf{k}}.r_i)a^\dagger(\mathrm{\mathbf{k}})\\
a_i &=\sum_{\mathrm{k}\in B}\exp(i\mathrm{\mathbf{k}}.r_i)a(\mathrm{\mathbf{k}})
\end{align}
\label{e:operators}
\end{subequations}

Substituting Eq. (\ref{e:operators}) in Eq. (\ref{e:SW2}), we obtain the Hamiltonian,
\begin{equation}	
	H(k)=H_0(k)\left(1-\frac{\langle n \rangle}{2s}\right).
	\label{e:SW3}
\end{equation}
The eigenvalue of Eq. (\ref{e:SW3}) is the excitation energy $E_l(\mathrm{\mathbf{k}},T)$ for the $l^\mathrm{th}$ band.
Here, $E_l(\mathrm{\mathbf{k}},T)$ is temperature dependent because ${\langle n (T) \rangle}$ follows Bose-Einstein statistics, 
\begin{equation}
\langle n (T) \rangle=\int_{\BZ}\sum_l\frac{d^2\mathrm{\mathbf{k}}}{\exp(\frac{E_l(\mathrm{\mathbf{k}},T)}{\kBT})-1}.
\end{equation}    

To calculate the Curie temperature, the starting spin-configuration is considered to be pointing in the $z$ direction. 
The magnetization is defined as $S=\langle \hat{S}^z \rangle$, with $\langle \hat{S}^z \rangle=S-\langle a^\dagger a \rangle$ (Holstein-Primakoff transformation).
As defined earlier, $\langle a^\dagger a \rangle=\langle n (T) \rangle$, leading to the equation for magnetization as a function of temperature, 
\begin{equation}
	S(T)=S-\frac{1}{N}\int_{\BZ}\sum_l\frac{d^2\mathrm{\mathbf{k}}}{\exp(\frac{E_l(\mathrm{\mathbf{k}},T)}{\kBT})-1}.
	\label{e:self}
\end{equation}
Here, $N$ is the number of atoms in the unit-cell, $l$ is the band index, and $S$ is the initial magnetization.
Solving magnetization $S(T)$ and the energy eigenvalue $E_l(\mathrm{\mathbf{k}},T)$ self-consistently, we obtain the temperature dependent magnetization.
However, \Eq{e:self} diverges for $E_l(\mathrm{\mathbf{k}},T)=0$.
To avoid divergence, we define Curie temperature as the temperature at which $S(T)=S/2$.  

\subsection{\new{Green's function}}

Zubarev's double-time temperature-dependent Green's functions have been proven successful for three-dimensional ferromagnets in the past, and this over the entire temperature range~\cite{Zubarev60,Callen63}.
The technique explicitly accounts for the fact that spins obey bosonic commutation relations between distinct lattice sites and fermionic ones between different lattice sites.
First, one needs to derive an equation of motion for the Green's functions---this is an exact relation that derives from the Heisenberg equation of motion.
The Green's function equation of moton we derive is,
\begin{equation}
\omega G^\alpha_{ij}=\frac{1}{2\pi} \langle [\hat{S}^\alpha_i,\hat{S}^-_i]\rangle \delta_{ij}+\langle\langle[\hat{S}^\alpha_i,\hat{H}];\hat{S}^-_j\rangle\rangle.
\label{e:Green}
\end{equation}
Here, $i\omega$ is the excitation energy, $\hat{H}$ is the Heisenberg Hamiltonian, and $G^\alpha_{ij}=\langle\langle \hat{S}^\alpha_i;\hat{S}^-_j\rangle \rangle$ is the Green's function for the spin-operator $\hat{S}^\alpha_i$ with $\alpha \in \{+,-,z\}$. 
With some algebra it is easy to see that the higher order Green's function $\langle\langle[\hat{S}^\alpha_i,\hat{H}];\hat{S}^-_j\rangle\rangle$ reduces to $\langle\langle\hat{S}^\alpha_i\hat{S}^\beta_j;\hat{S}^-_j\rangle\rangle$, where $\beta \in \{+,-,z\}$.
To allow for a solution of the Green's function equation of motion (Eq. (\ref{e:Green}), higher order Green's functions $\langle\langle\hat{S}^\alpha_i\hat{S}^\beta_j;\hat{S}^-_j\rangle\rangle$ are decoupled in terms of lower order Green's function $G^\alpha_{ij}$ using the Tyablikov decoupling approximation~\cite{Tyablikov59} (which gives the same results as Englerts random phase approximation combined with the appropriate form of the fluctuation-dissipation theorem~\cite{Englert60}). The Tyablikov decoupling scheme decouples higher order terms using,
\begin{equation}
\begin{matrix}
\langle\langle\hat{S}^\alpha_i\hat{S}^\beta_l;\hat{S}^-_j\rangle\rangle  &\rightarrow \langle \hat{S}^\beta_l\rangle \langle\langle\hat{S}^\alpha_i;\hat{S}^-_j\rangle\rangle+\langle \hat{S}^\alpha_i\rangle \langle\langle\hat{S}^\beta_l;\hat{S}^-_j\rangle\rangle \\
  &=\langle \hat{S}^\beta_l\rangle G^\alpha_{ij}+\langle \hat{S}^\alpha_i\rangle G^\beta_{lj}.
\label{e:Tyablikov2}
\end{matrix}
\end{equation}

We then define the homogenous magnetization $M=\langle\hat{S}^z\rangle$ and write the Green's function in the reciprocal space,
\begin{equation}
G^\alpha_{ij}=\frac{1}{N}\sum_{\mathrm{k} \in B}\exp(\mathrm{i\mathrm{\mathbf{k}}}.(r_i-r_j))G^\alpha(\mathbf{k})
\label{e:Green_FT}
\end{equation}

Combining Eq. (\ref{e:Green}), Eq. (\ref{e:Tyablikov2}), and Eq. (\ref{e:Green_FT}), we obtain a matrix equation,
\begin{equation}
(\omega \rm{\mathbf{I}}-\mathbf{\Gamma(k)}) \mathbf{G(k)}=\mathbf{A}.
\label{e:Green2}
\end{equation}

$\mathbf{I}$ is an identity matrix of size $3\times3$, $\mathbf{G(k)}$ comprises of three Green's functions $\{G^+,G^-,G^z\}$. 
For details on building matrices $\mathbf{\Gamma (\rm k)}$ and $\mathbf{\rm A}$, the interested reader may refer to Vanherck et.al ~\cite{Vanherck18,Vanherck20}.

The Green's function in Eq. (\ref{e:Green2}) is solved self-consistently with the homogenous magnetization $M$ for each temperature. 
The temperature at which the homogenous magnetization $M$ becomes $0$ is referred to as the Curie temperature. 
However, in the absence of an external field, taking the limit of small magnetization (close to the ferromagnetic transition temperature) yields an explicit expression for the Curie temperature.
For effective easy-axis anisotropies $\sum_j J_{ij} \Delta_{ij} \leqslant 0$, the Curie temperature vanishes in accordance to the Mermin-Wagner theorem \cite{Mermin66}.
On the other hand, $\TCurie$ for effective easy-axis anisotropies can be written as~\cite{Vanherck18,Vanherck20}
\begin{equation}
	\kB\TCurie = \frac{S (S+1)}{3 \Phi_\Curie},
	\quad
	\Phi_{\Curie}
	=
		\Over{\RecPrimCellVol}
		\dint{\BZ}{}{
			\phi_{\Curie}(\vk)
		}{\vk}.
\end{equation}
The integrand is
\begin{equation}
	\phi_{\Curie}(\vk)
	=
		\frac{1}{T - f_{\Equal} - \norm{f_{\Other}}}.
\end{equation}
Let $\sum_{n}$ represent a sum over all n\nth{th} neighbours, \ie NN, NNN, and $\sum_{n,\Equal}$ and $\sum_{n,\Other}$ the same sum but restricted to atoms located on the same (Equal) or the other sublattice.
We define
\begin{equation}
	T=\sum_{n} J_n (1+\Delta_n), f_{\Equal} = \sum_{n,\Equal} F_n\text{, and }f_{\Other} = \sum_{n,\Other} F_n.
\end{equation}
$T$ is a measure for the total anisotropic exchange interaction.
$f_{\Equal}$ and $f_{\Other}$ are defined in terms of $F_n = J_n (1-\Delta_n)\sum_{p}\exp(\ii \vk \bcdot \vec{r}_p) \delta_{np}$.
Here, $\delta_{np}=1$ for $n=p$.

\new{As compared to the renormalized spin-wave theory, not only the excitation energies but also the effective density of states is renormalized by the magnetization, yielding for a better description over the entire temperature range.
The major difference between the Greens function and the RNSW lies at the level the decoupling is performed. 
The Tyablikov decoupling~\cite{Tyablikov59} is performed at the level of spin-operators. 
Whereas the Hartree-Fock decoupling is performed at the level of Bosonized spin-waves.
}

\subsection{Monte-Carlo}

For both the Ising and the Monte-Carlo with anisotropy, \Eq{e:Heisenberg} is treated as a classical equation with spin-operators ($\vec{\hat{S}}$) becoming spin-vectors ($\mathbf{S}$).
We use the Metropolis-algorithm to simulate the phase change of the classical Heisenberg Hamiltonian~\cite{sample9}.

For the Metropolis sampling of Ising Monte-Carlo, the spin-vectors become scalars and are fixed to take values $S\in\{-S_{\mathrm{max}},S_{\mathrm{max}}\}$.
Whereas for the Monte-Carlo with anisotropy, the spin-vectors are sampled using a spherical sampling scheme~\cite{my-paper}.
From the Metropolis algorithm, we obtain the magnetic susceptibility and specific-heat as a function of temperature.
We obtain the Curie temperature from the peak of specific-heat or susceptibility as they both coincide for easy-axis ferromagnets.

\subsection{DFT calculations}
\new{The $J$, $\Delta$, and $S$ were directly obtained from the C2DB database and were fed to the analytical formulas. 
However, we calculated the long-range $J$-parameters for Cr-compounds and $\rm{Fe_2F_2}$, using the method developed in Ref.~\onlinecite{my-paper}, which uses non-collinear DFT calculations. 
All the ab-initio DFT calculations reported in this work were performed using the Vienna ab-initio simulation package (VASP)~\cite{sample3,sample5}.
The ground state self-consistent field (SCF) calculations were performed using a projector-augmented wave (PAW) potential~\cite{sample3} with a generalized-gradient approximation as proposed by Perdew-Burke-Ernzerhof (PBE)~\cite{sample4}.
We have used a kinetic energy cut-off of 400 eV for our DFT calculations.
The Brillouin zones were sampled using a $\Gamma$-centred $k$-point mesh of size $5\times5\times1$ points for $2\times1\times1$ supercells.
The Cr-compound supercells were relaxed until the force on each of the ions was below $10\,\mathrm{meV}/\mathrm{\AA}$.
The energy convergence criterion for the subsequent SCF calculations was set to $10^{-4}\,\mathrm{eV}$.
The C2DB parameters and our own DFT calculations for the Cr-compounds and $\rm {Fe_2F_2}$ showed a difference less than 20\% in the $J$-parameters.
}

\section{Acknowledgements}
The project or effort depicted was or is sponsored by the Department of Defense, Defense Threat Reduction Agency.
The content of the information does not necessarily reflect the position or the policy of the federal government, and no official endorsement should be inferred.

This work was supported by imec's Industrial Affiliation Program.

\section*{References}
\bibliography{bib}
\end{document}